\documentclass[12pt,a4paper]{article}
\pdfoutput=1

\usepackage{xspace}
\usepackage{latexsym}

\usepackage{ifthen}
\newboolean{commentsaon}         %
\setboolean{commentsaon}{true}
  \setboolean{commentsaon}{false}  %

\newboolean{commentson} %
\setboolean{commentson}{true}

\newcommand{\comment}[1]
{\ifthenelse{\boolean{commentson}\AND\boolean{commentsaon}}
   {{\par\noindent\mbox{}{\footnotesize\blue[ *** #1 ]\par}\noindent\par}}{}}

\newcommand{\commenta}[1]
{\ifthenelse{\boolean{commentsaon}}
   {{\par\noindent\mbox{}{\small\color[rgb]{0, .5, 0}[ *** #1 ]\par}\noindent\par}}{}}

\usepackage[yyyymmdd]{datetime}

\newcommand{\myhalfmagnification}
{0.028}

\newcommand{\extraheight} {.06}
\usepackage{color}
\newcommand\blue     {\color{blue}}

\usepackage{hyperref}

\newtheorem{theorem}{Theorem}
\newtheorem{corollary}[theorem]{Corollary}
\newtheorem{lemma}[theorem]{Lemma}
\newtheorem{proposition}[theorem]{Proposition}
\newtheorem{definition}[theorem]{Definition}
\newcommand*{\seq}[2][n]  {{#2_{1}, \allowbreak \ldots, \allowbreak #2_{#1}}}

\newcommand*{\Var}{{\ensuremath{\it Var}}\xspace}

\newcommand*{\myunderscore}{\mbox{\tt\symbol{95}}}

\newcommand*{\nqueens}{{\sc nqueens}\xspace}

\title{{SLD}-resolution without occur-check, \\ an example%
        {\ifthenelse{\boolean{commentson}\AND\boolean{commentsaon}}
           {\blue\quad  \normalsize[(version 2.1)]}{}%
        }%
}%

\author{W{\l}odzimierz Drabent
\\[1ex]
\normalsize
\begin{tabular}{c}
    Institute of Computer Science,  Polish Academy of Sciences
\\\normalsize
   IDA, Link\"oping University, Sweden%
\\
{\tt\normalsize\small
      drabent\,{\it at}\/\,ipipan\,{\it dot}\/\,waw\,{\it dot}\/\,pl}
\end{tabular}
}

\begin{document}
\maketitle

\begin{abstract}
\noindent
We prove that the occur-check is not needed for a certain definite clause
logic program, independently from the selection rule.
First we prove that the program is occur-check free.
Then we consider a more general class of queries, under which the program is
not occur-check free;
however we show that it will be correctly executed under Prolog
without occur-check.

The main result of this report states that
the occur-check may be skipped
for the cases in which
a {\em single} run of 
a standard nondeterministic unification algorithm does not fail due to the
occur-check. 
The usual approaches are based on the notion of NSTO (not subject to
occur-check), which considers all the runs.
To formulate the result, it was necessary to
introduce an abstraction of a ``unification'' algorithm 
without the occur-check.

\end{abstract}

\section{Introduction}
Programming language Prolog implements SLD-resolution, employing an unsound
implementation of unification without the occur-check.
In practice this creates no problems.  Programmers know that they do not need
to care about it, unless they deal with something like difference lists.
Then one should be careful at tasks like checking a difference list for
emptiness.%
\footnote{
  In the important Prolog textbook by Sterling and Shapiro
  \cite{Sterling-Shapiro-shorter}, the occur-check is 
mentioned (in the context of actual programs) only when dealing 
with checking a difference list for emptiness (p.\,298--300).

}
As the occur-check is so unimportant in practice, 
one would expect that it should be easy to establish formally 
that it can safely be skipped.
For instance, two sufficient conditions 
for occur-check freeness are presented in the
textbook  \cite{apt-prolog}.
They are however applicable only for LD-resolution (SLD-resolution under
the Prolog selection rule).  Moreover they turn out inapplicable for some
simple programs, like
the $n$ queens program by {Fr{\"u}hwirth}
\cite{Fruehwirth91}.
They are based on the notion of NSTO
(not subject to occur-check) \cite{DBLP:conf/slp/DeransartFT91,apt-prolog},
 which requires that the occur-check
never succeeds
within each possible run of a non-deterministic unification algorithm,
for the input of interest.

In this paper
 we prove that for the $n$ queens program the occur-check is not needed.
First, in Section \ref{sec.occur-check.free}, we
 apply an ad hoc approach, and prove that the program with the
intended queries is occur-check free \cite{apt-prolog}, for any selection
rule.  (This means all the unifications in the SLD-derivations are NSTO).

Then, in Section \ref{sec.withoutOC}, we show that even when the unification
cases are not NSTO the occur-check may be not necessary.
We provide an appropriate sufficient condition.
It makes possible showing that the occur-check may be omitted also for some
programs/queries which are not occur-check free in the usual sense.
In particular, we show that 
the $n$ queens program may be executed
without the occur-check for a wider class of queries than those considered in 
Section \ref{sec.occur-check.free}.  Moreover the proof is simpler.

\enlargethispage{1pt plus 1.5ex}
\vspace{0pt minus 1.5ex}
\paragraph{Preliminaries.}

We use definitions, theorems etc from \cite{apt-prolog}
(and we do not repeat them here).
We refer to two unification algorithms, 
the Nondeterministic Robinson Algorithm
 \cite[p.\,28]{apt-prolog},
and
the (also nondeterministic)
Martelli-Montanari algorithm  (MMA) \cite[p.\,32]{apt-prolog}.
In particular we refer to the action numbers of MMA used there.
(The same algorithm is presented in \cite{nilsson.maluszynski.book},
however actions 5 and 6 are numbered, respectively 5b and 5a.)
As in \cite{apt-prolog}, the two usages of ``='' (as a syntactic symbol in
equations, and as the meta-language symbol for equality) are to be
distinguished by the context.

We need to generalize some definitions from \cite{apt-prolog}, 
in order not to be limited to LD-resolution.
We will say that {\em unification of $A$ and $H$ is available} in an
SLD-derivation (or SLD-tree)
for a program $P$,
if $A$ is the selected atom in a query of the derivation (tree), and $H$
is a standardized 
apart head of a clause from $P$,
such that $A$ and $H$ have the same predicate symbol.
(A more formal phrasing is {``equation set $\{A=H\}$ is available''}.)
If all the unifications available in an SLD-derivation are NSTO
 \cite[Def.\,7.1]{apt-prolog} then the derivation is {\em occur-check free}.
We say that a program $P$ with a query $Q$ is {\em occur-check free}
if, under a given selection rule, all SLD-derivations for $P$ with $Q$ 
are occur-check free.

By an expression we mean a term, an atom, a tuple of terms (or atoms),
or an equation.  An expression, a set of equations. or a substitution is 
{\em linear} when no variable occurs in it more than once.
As in Prolog, each occurrence of \myunderscore\ in an expression
(or an equation set, or a substitution)
will stand for a distinct variable.
Otherwise variable names begin with upper case letters.

\section{The program} 
We will deal with the core fragment of the $n$ queens program
\cite{Fruehwirth91}., we will call it \nqueens
\cite{drabent2019arxiv.nqueens}.

\pagebreak[3]
\vspace{\abovedisplayskip}
\noindent
\mbox{}\hfill%
\begin{minipage}[t]{.8\textwidth}
{\small %
\begin{verbatim}
    pqs(0,_,_,_).
    pqs(s(I),Cs,Us,[_|Ds]):-
            pqs(I,Cs,[_|Us],Ds),
            pq(s(I),Cs,Us,Ds).

    pq(I,[I|_],[I|_],[I|_]).
    pq(I,[_|Cs],[_|Us],[_|Ds]):-
            pq(I,Cs,Us,Ds).
\end{verbatim}
}
\end{minipage}%
\hfill
 \begin{minipage}[t]{.035\textwidth}
\raggedleft
\small%
(\refstepcounter{equation}\theequation\label{clause1})
\\ \ \\
       \refstepcounter{equation}%
      (\theequation\label{clause2})%
\\ \ \\ \ \\ 
       \refstepcounter{equation}%
      (\theequation\label{clause3})
\\[1.5ex]
       \refstepcounter{equation}%
        {(\theequation\label{clause4})}
\end{minipage}%
\vspace{\belowdisplayskip}
\\
The typical initial query is 
$Q_{\rm in}={\it p q s}(n,q_0,\myunderscore,\myunderscore)$, where
$q_0$ is a list of distinct variables, and $n$ a natural number represented
as $s^i(0)$.
We only mention that the sufficient conditions from \cite{apt-prolog} and
\cite{AptL95.delays} for occur-check freeness are inapplicable for \nqueens
with such query.
Note that:
\begin{lemma}
\label{lemma.1ground}
   If the first argument of any predicate symbol in an initial query
  $Q_0$ is ground, then
  in any query of any SLD-derivation of \nqueens starting with $Q_0$,
  the first argument of any predicate symbol is ground.
\end{lemma}

\section{Occur-check freeness of \nqueens}
\label{sec.occur-check.free}

Here we prove 
that \nqueens is occur-check free (for the intended initial queries).
The proof is based on the fact that all the atoms appearing in the queries of 
the SLD-derivations of
interest are linear.

\begin{lemma}
\label{lemma.linear.result}
Unification of a linear pair of atoms $A,H$
results in a linear atom.

If each of two pairs $A,H$ and $B,H$ is linear, and $\theta$ is an mgu of
$A$ and $H$ then $B\theta$ is linear.
\end{lemma}
\noindent
{\sc Proof }
The Nondeterministic Robinson Algorithm \cite[p.\,28]{apt-prolog},
 produces not only an mgu, but also
the result of the unification.  While unifying $A$ and $H$ it maintains a
current substitution $\theta$, and two instances $A\theta$ and  $H\theta$.
We show by induction  that 
at each step, each of $A\theta$,  $H\theta$ is linear, and if a variable $Y$
occurs in both of them then $Y$ does not occur within a disagreement pair
of $A\theta,\,H\theta$;
moreover, $B\theta$ is linear, and if $Y$ occurs in both  $B\theta$, $H\theta$
then $Y$ does not occur within a disagreement pair of  $A\theta,\,H\theta$.

{\small\footnotesize
Obviously this holds at the beginning, with
$\theta=\epsilon$. 

Assume that the property holds for 
$\theta$, $A\theta$,  $H\theta$, $B\theta$, 
The new current substitution is $\theta'= \theta\{X/t\}$, where $X,t$ (or $t,X$) is a
disagreement pair. 
So $X$, and each variable from $\Var(t)$ occurs exactly once in 
$A\theta,H\theta$.  
Moreover, $t$ is a subterm of that atom from $A\theta,H\theta$
in which $X$ does not occur.
So replacing $X$ by $t$ in $A\theta$ or in $H\theta$ results in a linear atom.
(The other atom is unchanged by applying $\{X/t\}$ to it.)
Thus each $A\theta'$ and $H\theta'$ is linear.

Each variable of $t$ occurs once in $A\theta'$ and once in $H\theta'$,
but this is not within any disagreement pair of $A\theta'$ and $H\theta'$.
If a variable  $Y$ occurs in both $A\theta'$ and $H\theta'$
then it occurs in $A\theta$ and in $H\theta$, or in $t$.
In both cases it does not occur within a disagreement pair
of $A\theta'$ and $H\theta'$.

\smallskip
Let us now consider $B\theta'$.
If $X$ does not occur in $B\theta$ then
$B\theta'$ is $B\theta$, hence linear.
Otherwise $B\theta'$ is obtained by replacing the single occurrence of $X$ 
in $B\theta$ by a linear term $t$.
Variable $X$ does not occur in $H\theta$ (otherwise $X$ does not occur in a
disagreement pair of $A\theta$ and $H\theta$, contradiction).
So $t$ is a subterm of $H\theta$.  As $t$ occurs within a disagreement pair
of    $A\theta,H\theta$, 
any variable of $t$ does not occur in $B\theta$, hence it occurs exactly once
in  $B\theta'$.  Now 
$\Var(B\theta')\setminus \Var(t) = \Var(B\theta)\setminus\{X\}$,
each of these variables occurs exactly once in $B\theta$ and thus exactly
once in $B\theta'$. 
Hence $B\theta'$ is linear.

Consider a variable $Y$ that occurs in both $B\theta'$ and $H\theta'$.
Similarly to the previous case, $Y$ occurs in both $B\theta$ and $H\theta$,
or in $t$.  In both cases it does not occur within a disagreement pair
of $A\theta'$ and $H\theta'$.  {\normalsize}
\par
}

Thus, if an mgu $\theta$ of $A,H$ is obtained, then $A\theta$ is linear, 
and so is $B\theta$.
Any other mgu of $A,H$ is $\theta\gamma$ (where $\gamma$ is a renaming).
Hence both $A\theta\gamma$ and $B\theta\gamma$ are each linear.
 $\Box$

\begin{proposition}
\label{lemma.each.linear}
\label{propo.each.linear}

Let $Q_0 = {\it p q s}(n, t_1,t_2,t_3)$, where $n$ is ground, be linear.
In any SLD-derivation for \nqueens and $Q_0$, each atom in each query is
linear.
\end{proposition}

\noindent
{\sc Proof} 
by induction. \ The first query $Q_0$ of the derivation consists of one linear
atom. 
Remember that
the first argument of each atom in the derivation is ground
(by Lemma~\ref{lemma.1ground}).

Assume that each atom in a query $Q_i$ is linear.
Let  $Q_{i+1}$ be the next query in the SLD-derivation.
If the clause applied in the resolution step has a linear head
then, by Lemma \ref{lemma.linear.result},
each atom of $Q_{i+1}$ is linear (as such atom is obtained by applying the
mgu to a linear atom from the body of a clause or from $Q_i$).
This concerns clauses  (\ref{clause1}), (\ref{clause2}) and (\ref{clause4}).

For clause  (\ref{clause3}), the selected atom from $Q_i$ is 
$A={\it p q}(m, u_1,u_2,u_3)$, where $m$ is ground.
It is unified with a standardized apart head
$H={\it p q}(I,[I|\myunderscore],[I|\myunderscore],[I|\myunderscore])$.
$A$ and $H$ are unifiable iff
$E=\{
I=m,\,u_1=[I|\myunderscore],\linebreak[3]
u_2=[I|\myunderscore],\linebreak[3]\, u_3=[I|\myunderscore]
\,\}$
is unifiable. 
By Lemma 2.24 (Iteration) of \cite{apt-prolog},
$E$ is unifiable iff
$E_2= \{ u_1=[m|\myunderscore],\linebreak[3]
u_2=[m|\myunderscore],\linebreak[3]\, u_3=[m|\myunderscore] \,\}$ is
(as $\{I/m\}$ is an mgu of $I=m$, and
 $u_i\{I/m\}=u_i$ and $[I|\myunderscore]\{I/m\}=[m|\myunderscore]$,
for $i=1,2,3$).
Moreover, an mgu of $E$ is $\{I/m\}\theta_2$, where $\theta_2$ is an 
mgu of $E_2$.   Note that $\theta_2$ is an mgu of
$A$ and $H\{I/m\}$.  As  $H\{I/m\}$ is linear, Lemma \ref{lemma.linear.result}
applies, and we obtain that each atom of $Q_{i+1}$ is linear.  $\Box$
{\sloppy\par}

\medskip

Note that the intended initial queries ($Q_{\rm in}$) are of the form
considered in Proposition \ref{propo.each.linear}.
As unification of two variable disjoint atoms, one of which is linear, is
not subject to occur-check
(by  Lemma 7.5 (NSTO) from \cite{apt-prolog}), it immediately follows:
\begin{corollary}
\label{corollary1}
\hspace{0pt plus .5em}
Program \nqueens with
a linear query
$Q_0 = {\it p q s}(n, t_1,t_2,t_3)$, where $n$ is ground,  is
occur-check free %
under any selection rule.
\end{corollary}

\section{NSTO is a too strong requirement}
\label{sec.withoutOC}
NSTO states that each execution of MMA for a given equation set $E$ does not
require the occur-check.  In this section
we show a weaker condition under which the occur-check is not needed.
As a motivating example,
we first show that
if MMA follows a certain left to right strategy,
then the occur-check is not needed for
 \nqueens 
with some queries for which the program is not occur-check free.

Then we formalize a notion of MMA without the occur-check.
This is a basis to formulate the main result of this section:
if some run of MMA does not involve 
the occur-check then each run of MMA without the occur-check produces a
correct result. 
Based on this, we show 
that \nqueens can be correctly executed without the occur-check
also for some queries for which  \nqueens is not occur-check free.

Now we show that the requirement of linearity from the previous section
is not needed for executing \nqueens without the occur-check.
Assume that
algorithm MMA selects equations from left to right.  Using the notation 
from the proof of Proposition \ref{propo.each.linear}, let us consider 
unification of a possibly not linear atom 
$A={\it p q}(m, u_1,u_2,u_3)$ (where $m$ is ground)
with a standardized apart head
$H$ of clause 
(\ref{clause3}). 
After the first two steps we obtain equation set $E$ (as in the proof).
Then $\{I/m\}$
 is applied to $E\setminus\{I=m\}$.  This results in 
$\{I=m\}\cup E_2$, which is NSTO (by  Lemma 7.5 from \cite{apt-prolog},
as the tuple $m, [m|\myunderscore],[m|\myunderscore],[m|\myunderscore]$ of the right hand sides of the
equations of $E_2$
is linear).
As all the other clause heads are linear, 
all SLD-derivations starting from a query 
$Q_0' = {\it p q s}(n, t_1,t_2,t_3)$ where $n$ is ground (and $t_1,t_2,t_3$
are arbitrary)
do not require the occur-check, provided MMA works as described above.
So we showed the following property.

\begin{proposition}
\label{propo.run.exists}
\label{lemma.run.exists}
Consider a query  $Q_0' = {\it p q s}(n, t_1,t_2,t_3)$, where $n$ is ground.
    There exists a strategy of selecting actions in MMA, so that
    action (6) (halt on the occur-check) is not performed
    in any unification available
     in any SLD-derivation for \nqueens with $Q_0'$.
\end{proposition}
However the program with such query is not occur-check free; 
under some other selection
of actions in MMA action (6) may be performed.
Consider for instance
$A={\it p q}(m, L, [L|\myunderscore],\myunderscore)$ (with a ground $m$).
A run of MMA on $A$ and $H$, 
as described above, does not employ the occur-check and halts 
with failure due to action (2).
But some other run, which delays applying $\{I/m\}$, halts on the occur-check
(action (6)).
We are going to show that even in such cases
MMA without the occur-check would produce correct results (by failing on
action (2), after having produced infinite terms).

Tests under SICStus and SWI-Prolog confirm that Prolog behaves in this way.
Prolog unification without the occur-check
applied to $A$ and $H$
correctly results in failure.  The same happens when the arguments 
of $A$ and $H$ are permuted
(by the same permutation);
 we can assume that this changes
the order of elementary actions of the Prolog unification.
We may suppose that in the latter case infinite terms are created.
A particular order of actions of the Prolog algorithm 
on
 $A$ and $H$ can be simulated by a Prolog query
{\small\footnotesize
  \verb:(L,[L|_],_)=([I|_],[I|_],[I|_]),: \verb:s(0)=I:} %
(which fails).
Posing only the first conjunct of this query shows, as expected, that 
the unsound Prolog algorithm ``unifies'' the two triples, producing 
a ``unifier'' containing infinite terms.

\paragraph{MMA without the occur-check. }
Here we introduce a version of the Martelli-Montanari algorithm without the
occur-check.   We will call it MMA$\!^-$.
It is a generic algorithm.  To make sense, it has to be executed under a
particular strategy of selecting equations and actions to be performed.
This is needed, among others,to assure termination.
It is not clear which (unsound) unification algorithms are used by actual
Prolog implementations.  However we conjecture that the runs of such 
algorithms are sufficiently well abstracted by some runs of MMA$\!^-$.

MMA$\!^-$ differs from MMA  \cite[p.\,32]{apt-prolog} by
\[
\begin{tabular}{l@{ }l}
(a) & removing action (6),
\\
(b) &
\parbox[t]
{.75\textwidth}
{ replacing condition $X\not\in\Var(t)$ in action (5) by 
$X\not=t$,
}
\\
(c) & 
adding action \\
\multicolumn{2}{l}
{%
  \begin{tabular}{l@{ }l}
    (5') &
    \parbox[t]{.74\textwidth}{
    $X=t$ where $X\not=t$, and
    variable $X$ occurs elsewhere 
      \qquad$\longrightarrow$ \\
      replace {\em some} occurrences of  $X$ by $t$ in the other equations,
    }
  \end{tabular}%
}
\\[3.5ex]

(d) &
\parbox[t]
{.76\textwidth}
{ 
 adding a requirement that action (5') can be performed 
only if  (5) has been previously performed
on $X=t'$, for some $t'$.
} %
\end{tabular}
\]
\noindent
So action (5') is a generalization of (5)
(which replaces {\em all} the occurrences of~$X$).
Due to requirement (d), action (5') is performed only if
the equation set $E$ dealt with is not unifiable 
(as if after the previous usage of (5) for %
 $X=t'$
the variable $X$ occurs more than once, then  $X\in\Var(t')$).
So, in a sense, (5') is applied only if we deal with equations representing
infinite terms.
Equation $X=t$ can be understood as describing such term,
and the occurrences of $X$ in the other equations as references to the term.
(Immediately after (5) each such occurrence occurs within a subterm $t$.)

It remains to describe the results of MMA$\!^-$.

\begin{definition}\rm
  Consider a set of equations $ E=\{X_1{=}t_1,\ldots,X_n{=}t_n\}$, 
  and a relation $\succ_E$ on variables $\{\seq X\}$
  defined by $X_i\succ_E X_k$ iff
  $X_k\in\Var(t_i)$ (for $i,k\in\{1,\ldots,n\}$).
  Let $\succ_E^+$ be the transitive closure of $\succ_E$.

  $E$ is  {\em semi-solved} if $\seq X$ are distinct, 
  no $t_i$ is $X_i$, and
  if $X_i$ occurs in $t_k$ then  $X_i\succ_E^+ X_i$,
  for any  $i,k\in\{1,\ldots,n\}$.
\end{definition}
So e.g.\ 
$\{ X{=}f(X),\,Y{=}X \}$ and
$E_1=\{ X{=}f(Y), Y{=}f(X) \}$ are semi-solved, but 
$E_1\cup \{X=a\}$ and 
$\{ X{=}a, Y{=}f(X) \}$ are not.
Note that $E$ is {\em solved} \cite{apt-prolog} when 
$\seq X$ are distinct and $\succ_E$ is empty.
Solved equation sets
are results of MMA when it
successfully terminates.  Semi-solved ones play the same role for MMA$\!^-$.

Now we are ready to describe termination of MMA$\!^-$.
The algorithm may terminate with failure or with success.
{\em Termination with failure} is explicitly caused by action (2).
{\em Termination with success} happens when the
current equation set is semi-solved.
Note that actions (5) or (5') may be applicable to such equation set,
in such case the nondeterministic MMA$\!^-$ may terminate or continue. 
Note also that if $E$ is not semi-solved
then some action of  MMA$\!^-$
can be applied to $E$.  Thus
a finite run of MMA$\!^-$ either terminates with failure, or with success.

We conjecture that it is sufficient to (a) not apply action (5) to $X=t$
if it has been previously performed on some $X=t'$, and
(b) apply action (5')
only to occurrences of $X$ that are left hand sides of equations of the form 
$X=t'$, moreover where $|t|\leq|t'|$.%
\footnote{ So (5') transforms $X=t'$ into $t=t'$, and the size of the/a larger
  term of an equation is preserved.
}
  (Here $|s|$ stands for the size of a term $s$, i.e.\ the
  number of occurrences in $s$ of variables and function symbols, including
  constants.)

\smallskip

{\footnotesize
In some older Prolog systems the (unsound) implementation of unification 
did not terminate in some cases \cite{MarriottS89}.
It seems that it always terminates in the current systems.
Also, MMA$\!^-$  may not terminate.
We conjecture that it terminates with the restrictions from the previous
paragraph on applying actions (5) and (5').
We expect that a termination proof similar to that of 
\cite{Colmerauer1982} can be constructed.
E.g.\ take  $E=\{X{=}f(X),\,X{=}f(f(X))\}$.
Applying (5) to $X{=}f(X)$  and then (1) to $f(X){=}f^3(X)$
results in $E$ again.
Now there are two ways of applying
(5') to a left hand side occurrence of $X$.  
(5') with $X=f(X)$ selected and then action (1) applied to $f(X){=}f(f(X))$ 
results in  $\{X{=}f(X)\}$, which is semi-solved.
(5') with $X=f(f(X))$ selected and then action (1) applied to $f(f(X)){=}f(X)$ 
produces $E$ with the first equation reversed.
This may be extended to an infinite loop.
Note that the latter case violates one of the restrictions above;
$t=f(f(X))$ replaces the left hand side of $X=f(X)$, however 
$|t|>|f(X)|$.
\par
} %

\paragraph
{Correct runs of MMA$\!^-$.} 
By an {\em occur-check free run} of MMA we mean a run of MMA
in which action (6) is not performed.   
Note that an occur-check free run of MMA
is also a run of MMA$\!^-$.  
Conversely, each run of MMA$\!^-$ in which whenever action (5) is applied to
an equation
$X=t$ then $X\not\in\Var(t)$ is an occur-check free run of MMA
(as in such run of MMA$\!^-$ action (5') cannot be performed).
So if $E$ is NSTO then each run of MMA$\!^-$ for $E$ is a run of MMA.
The latter properties were the reason to introduce requirement (d)
in MMA$\!^-$.

We will say that a run of MMA$^-$ on an equation set $E$ is {\em correct},
if it produces the right result
 i.e.\ the run halts with failure
if $E$ is not unifiable, and produces an mgu of $E$ otherwise.
(Formally, producing an mgu means obtaining a solved equation set $E'$,
such $E'$ uniquely represents the mgu.)
\begin{theorem}
\label{lemma.MMAminus}
\label{propo.MMAminus}
\label{th.MMAminus}
Consider an equation set $E$.  Assume that there exists an occur-check free
run of MMA on $E$.  Then if a run of MMA$^-$ for $E$ terminates then it is
correct. 
\end{theorem}

For a proof we first introduce a few notions.  We will consider possibly
infinite terms ({\em i-terms}) over the given alphabet of function symbols,
including constants.  
The corresponding generalization of the
notion of substitution is called {\em i-substitution}.  We will follow the
ideas from 
the study of MMA in \cite{apt-prolog}, but we will employ i-terms.
An i-substitution $\theta$ is an 
{\em i-solution} of an equation $t=u$ if
$t\theta$ and $u\theta$ are the same expression;
$\theta$ is a solution of a set $E$ of equations, if 
$\theta$ is a solution of each equation from $E$.
Two sets of equations are {\em i-equivalent} if they have the same set of 
i-solutions.

\begin{lemma}
\label{lemma.action.MMAminus}

  Each action of MMA$^-$ replaces an equation set by an i-equivalent one.
\end{lemma}

\noindent
{\sc Proof.}
For actions (3), (4) the claim is obvious.  The proof for (1) is as in the
case of MMA \cite[Claim 2, p.\,34]{apt-prolog}.
Action (5) or (5') replaces an equation set
$E=E_X\cup E_1$ by $E'=E_X\cup E_2$, where $E_X=\{X\mathop{=}t\}$.
Consider a solution $\theta$ of $E_X$.
So $X\theta, t\theta$ are the same term. As $E_2$ results from $E_1$
by replacing some occurrences of $X$ by $t$, $E_1\theta$ and $E_2\theta$ are
the same.  
Thus $\theta$ is a solution of $E$ iff  $\theta$ is a solution of $E'$.
$\Box$

\medskip\noindent
{\sc Proof}
 (of Theorem \ref{th.MMAminus}) \
Let $R_1$ be an occur-check free run of MMA on $E$, hence of MMA$^-$ on $E$.
Let $R_2$ be an arbitrary run of MMA$^-$ on $E$.
Let $S$ be the set of the i-solutions of $E$, and thus of every equation set
$E'$ appearing in $R_1,R_2$ (by Lemma \ref{lemma.action.MMAminus}).

If $R_1$ is successful, then $E$ is unifiable, thus $S$ contains unifiers of
$E$, and thus of any $E'$  appearing in $R_1,R_2$.
Hence no equation in any equation set of $R_2$ is of the form $X=t$, where
$X$ occurs in $t$ and $X\neq t$.  Thus if action (5) is performed in $R_2$
for some $X=t$, then after this step each equation set contains exactly one
occurrence of $X$.  Thus action (5') is never performed in $R_2$.
Hence $R_2$ is a run of MMA, so it produces an mgu of $E$.

Assume that $R_1$ halts with failure.  This must be due to the last step
performing action (2).  So the last equation set contains $f(\ldots)=g(\ldots)$,
hence $S$ is empty.  So no semi-solved equation set appears in $R_2$
(as each such set has an i-solution).  Thus $R_2$ does not terminate with
success.  If it terminates then it terminates with failure
$\Box$

\medskip
From Proposition \ref{propo.run.exists} and Theorem \ref{th.MMAminus}
we immediately obtain:

\begin{corollary}
Assume that MMA$^-$ is executed under a strategy which results 
in termination of MMA$^-$.
Consider a query  $Q_0' = {\it p q s}(n, t_1,t_2,t_3)$, where $n$ is ground.
MMA$^-$ is correct for any available unification in any
SLD-derivation for \nqueens and  $Q_0'$,

\end{corollary}
\label{corollary2}
In other words, the program with such queries will be correctly
executed by Prolog, despite %
the lack of occur-check.
This also holds for any modification of the Prolog selection rule.

Note that the proof of the latter result is simpler than that of 
Corollary \ref{corollary1},
despite a more general class of queries dealt with.%
\footnote{
    The proof of Proposition~\ref{propo.run.exists} is substantially
    shorter than the union of those of
    Lemma \ref{lemma.linear.result} and Proposition~\ref{propo.each.linear}.  
    Even if we exclude Lemma \ref{lemma.linear.result} from the comparison,
    as it does not deal directly with the program,
    the proof of  Proposition \ref{propo.run.exists}
    seems simpler than that of Proposition \ref{lemma.each.linear} alone.
}
We just do not need to bother about
three arguments of the atoms in the queries.

\section{Conclusions}
This report proposes a generalization of the usual notions of NSTO (not
subject to occur-check) and occur-check free
\cite{DBLP:conf/slp/DeransartFT91,apt-prolog}.
They consider each execution of a nondeterministic unification algorithm.
We show that it is sufficient to consider a single execution.
We prove that if an execution of the algorithm for an input $E$
does not require the occur-check then
the occur-check may be dropped from any execution.
To formulate this property,
we described an abstract unification algorithm without the occur-check.

The report is focused on an example, a simple logic program for which 
the standard criteria for occur-check freeness from \cite{apt-prolog} and
\cite{AptL95.delays} are inapplicable.
SLD-derivations under arbitrary selection rules are considered.
First, a proof is presented that,
for a class of queries, the SLD-derivations of the program are occur-check free.
Then, based on the above-mentioned property,
we prove that the program can be correctly executed without the occur-check, 
for a wider class of initial queries.
The class also includes queries, for which the derivations are not
occur-check free.  
Moreover, the latter proof turns out to be simpler than the former one.

The main result of this work is extending behind NSTO the class of cases for
which unification without the occur-check works correctly.

\bibliographystyle{alpha}
\bibliography{bibpearl,bibmagic,bibs-s,bibshorter}

\end{document}